# A new giant species of placented worm and the mechanism by which onychophorans weave their nets (Onychophora: Peripatidae)

Bernal Morera-Brenes[1,2] & Julián Monge-Nájera[3]
1. Laboratorio de Genética Evolutiva, Escuela de Ciencias Biológicas, Universidad Nacional, Heredia, Costa Rica; bernal.morera@gmail.com
2. Centro de Investigaciones en Estructuras Microscópicas (CIEMIC), Universidad de Costa Rica, 2060 San José, Costa Rica.
3. Vicerrectoría de Investigación, Universidad Estatal a Distancia, San José, Costa Rica; julian.monge@ucr.ac.cr, julian-monge@gmail.com

**Abstract:** Onychophorans, or velvet worms, are poorly known and rare animals. Here we report the discovery of a new species that is also the largest onychophoran found so far, a 22cm long female from the Caribbean coastal forest of Costa Rica. Specimens were examined with Scanning Electron Microscopy; *Peripatus solorzanoi* **sp. nov.**, is diagnosed as follows: primary papillae convex and conical with rounded bases, with more than 18 scale ranks. Apical section large, spherical, with a basal diameter of at least 20 ranks. Apical piece with 6-7 scale ranks. Outer blade 1 principal tooth, 1 accessory tooth, 1 vestigial accessory tooth (formula: 1/1/1); inner blade 1 principal tooth, 1 accessory tooth, 1 rudimentary accessory tooth, 9 to 10 denticles (formula: 1/1/1/9-10). Accessory tooth blunt in both blades. Four pads in the fourth and fifth oncopods; 4th. pad arched. The previously unknown mechanism by which onychophorans weave their adhesive is simple: muscular action produces a swinging movement of the adhesive-spelling organs; as a result, the streams cross in mid air, weaving the net. Like all onychophorans, *P. solorzanoi* is a rare species: active protection of the habitat of the largest onychophoran ever described, is considered urgent. Rev. Biol. Trop. 58 (4): 1127-1142. Epub 2010 December 01.

**Key words:** new species, net weaving mechanism, largest onychophoran, fossil and extant velvet worms, Peripatidae, taxonomy, Costa Rica, *Peripatus*.

Onychophorans, or velvet worms, are small invertebrates that most biologists study in theory but due to their rarity, never see in the real life (Bouvier 1905, 1907, New 1995). There are two living families: the egg-laying Peripatopsidae occuring in Chile, South Africa and Oceania, and the Peripatidae that bears live young and occurs in the Neotropics, and in isolated tropical areas of Africa and Asia (Bouvier 1905, 1907, Ruhberg 1985).

Onychophorans are predators that hunt for small invertebrate prey that they capture with an adhesive net mainly composed of water and protein (Bouvier 1905, Read & Hughes 1987, Mora *et al*. 1996a, b). The phylum has been considered a landmark of the evolutionary process, sharing important features with both the annelids and the arthropods (Bouvier 1905, Ballard *et al*. 1992). In addition, the onychophorans are an ancient group that is known in fossil records from the mid Cambrian (Dzik & Krumbiegel 1989, Hou & Bergstrom 1995) and are regarded as the first animals that could raise their bodies from the substrate and walk (Monge-Nájera & Hou 2000). They originally were a marine taxon with varied body shapes, often protected by spiculae and armor, but the nearly 180 named extant species (Reid 1996, Trewick 1998, 2000) are all terrestrial, showing no spiculae or armor and presenting the same simple body plan (Monge-Nájera & Hou 2000).

Previously known onychophorans range in body length from 10mm through 15cm (Read

1988a, b; Ruhberg 1985) and it has been suggested that growth is limited by their tracheal respiratory system and by their lack of a hard skeleton (Monge-Nájera & Lourenço 1995). No formal studies were made, so far, dealing with size constraints caused by the hydraulic skeleton of Onychophora.

The scientific information about most species is limited to the species description and minimal collection data; mating behavior in the wild has not been properly documented (Ruhberg 1985, Tait & Briscoe 1995) and there are many taxonomical problems (Read 1988a, 1988b and Reid 1996). Here we describe the uncommonly sized *Peripatus solorzanoi* **sp. nov.** from the Caribbean coastal forest of Costa Rica, and provide additional information regarding its taxonomy, behaviour and conservation.

## MATERIALS AND METHODS

**Scanning Electron Microscopy:** Specimens were prepared for SEM according to standard procedures (Morera-Brenes & Monge-Nájera 1990).

**DNA isolation and sequencing:** DNA was extracted from tissue samples using the Wizard genomic DNA purification kit (Promega) according to the manufacturer's instructions. DNA concentration in aqueous solutions was measured spectrophotometrically and adjusted to 50ng/µl with TE. Polymerase chain reaction (PCR) amplification and sequencing conditions were previously described by Podsiadlowski *et al*. (2008). A 658 bp fragment (region) of the cytochrome c oxidase subunit I (COX1/COI) gene was amplified using the primer pair LCO1490/HCO2198 (Folmer *et al*. 1994). Sequences were obtained from two specimens and confirmed by double-check.

BLAST searches at GenBank provide useful mitochondrial cytochrome oxidase I (COI) gene sequences. Relevant species from every Onychophoran region were included at DNA analysis, including the few Peripatidae available: accession Ns.: *Epiperipatus biolleyi*, NC_009082-DQ666064 (Podsiadlowski *et al*. 2008), *Oroperipatus corradoi*, U62429 (Gleeson *et al*. 1998), and Peripatopsidae: *Metaperipatus inae*, NC_010961-EF624055 (Braband *et al*. 2010).

Sequences were analyzed using Geneious software (Biomatters). Default parameters settings for DNA global alignment with free end gaps (Blosum 62 cost matrix, gap opening 12, extension penalties 3) were used.

The protein-coding sequence of *P. solorzanoi* n.sp. was inferred by translation *in silico* using the invertebrate mitochondrial Genetic Code. The COI protein was compared to relevant taxa from the America's: Peripatidae: *Epiperipatus biolleyi* (ACCESSION ABF93293, Podsiadlowski *et al*. 2008), *Oroperipatus corradoi* (ACCESSION AAC95414, Gleeson *et al*. 1998), and Peripatopsidae: *Metaperipatus inae* (Mayer 2007; ACCESSION ABQ95564, Braband *et al*. 2010).

**Phylogenetic analysis:** This analysis was performed with Geneious Tree Builder. A genetic distance matrix was constructed using the Jukes-Cantor model, and the tree build method was based on the Neighbor Joining Algorithm (Biomatters 2009). A tree was drawn using the Tree View program (Page 1998).

Following Ruhberg (1985) we call the legs "oncopods". Whenever we use a term that is not used by all authors, we include the equivalent.

## SPECIES DESCRIPTION

**Species account:** *Peripatus solorzanoi*, n. sp.

**Holotype:** Female. Guayacán de Siquirres (Costa Rica, Limón Province, 10º02'58" N, 83º32'31" W, 400-500m.a.s.l.), 19 February 1996, Alejandro Solórzano. Museo de Zoología, Universidad de Costa Rica, San José (UCRMZ-59-01).

**Paratypes:** Four young that were born to holotype shortly after capture and other collected by Miguel Solano, Norberto Solano and Alejandro Solórzano, 13 August 2000. Museo de Zoología, Universidad de Costa Rica, San José (UCRMZ-60-01).

**Distribution:** Guayacán and Liverpool de Siquirres and Barbilla National Park, Limón Province, Costa Rica.

**Etymology:** *Peripatus solorzanoi* **sp. nov.** is dedicated to Costa Rican herpetologist Alejandro Solórzano, who discovered the species, in consideration of his extensive work on the Central American herpetofauna and for his frequent contribution of onychophoran specimens to the University of Costa Rica.

**Diagnosis:** The following combination of characters: *Dorsal primary papillae*: Convex and conical with rounded bases; more than 18 scale ranks. Basal piece separated from apical piece by a sligh constriction. Apical section dilated, spherical and symmetric, with a basal diameter of 20 ranks; 6 or 7 scale ranks on apical pieces. Sensory bristle central, thorn-shaped, straight or slightly curved with ornamented basis. Outer blade 1 tooth, 1 accessory tooth, 1 vestigial accessory tooth (formula: 1/1/1); inner blade 1 tooth, 1 accessory tooth, 1 rudimentary accessory tooth, 9 to 10 denticles (formula: 1/1/1/9-10). Accessory tooth blunt in both blades. Four pads in the fourth and fifth oncopods (oncopods="legs"). Nephridial tubercle free from third and fourth pads, in lateral posterior position. The 4th. pad is arched.

**Oncopods:** Males with 34 leg pairs (n=2) and females with 39-41 leg pairs (39, n=1; 40, n=2; 41, n=11). Total observed animals n=16.

**Foot papillae:** Three foot papillae, two anterior and one posterior (Fig. 1).

**Soles on foot (also known as *creeping pads*):** Four complete creeping pads, without presence of the vestigal fifth one in all oncopods.

**Nephridial tubercle:** Present in the 4th and 5th oncopod pairs, anteriorly displaced and opening between the 3rd and 4th creeping pads, free from the 3rd and indenting the proximal margin of 4th pad, which is crescent-shaped around it. Fourth creeping pad not divided by the nephridial tubercle (Figs. 8, 9).

**Integument:** *Structure of papillae*. Dorsal primary papillae convex and conical with rounded bases; without grooves parallel to the main body axis between them. Primary papillae of dorsal surface all of one type, conical, usually 5 to 12 accessory papillae between two of the larger ones. Basal pieces height of >18 scale ranks (Fig. 2). Accessory papillae oriented both to ridge and borders of each fold (Fig. 3).

Apical pieces symmetric, globular-shaped and dilated, with a basal diameter of about 20 ranks, and 6 to 7 scale ranks tall (Fig. 2, 3).

*Sensory bristles*. Thorn-shaped bristle, little developed, straight or slightly curved. Sensory

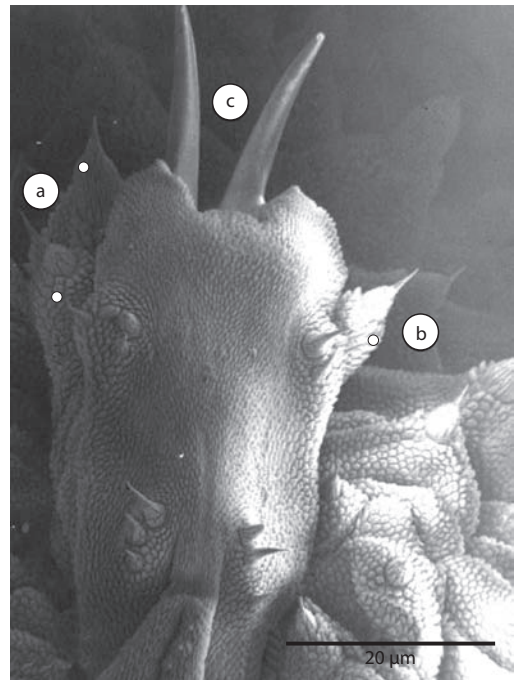

Fig. 1. Oncopod tip of *P. solorzanoi* in ventral view: foot papillae, (a) two anterior and (b) one posterior. (c) claws. Scale bar = 20 μm.

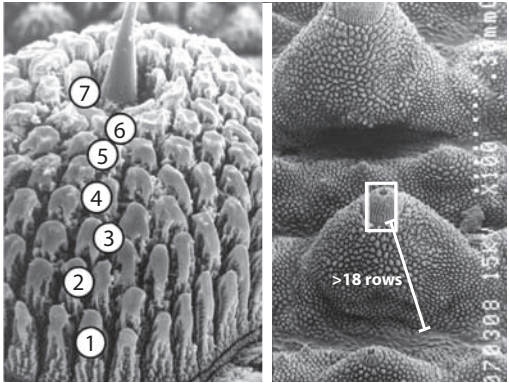

Fig. 2. Left. Apical section of the primary papillae, with large spherical shape. The apical piece has 7 rank scales. Right: primary papillae with a basal piece >18 scale ranks (rows of scales), from the base to the apical piece. Scale bar = 0.35 mm.

bristles placed centrally on the apical pieces and bearing an ornamented basis (Fig. 2).

*Plicae*. Dorsal integument with 12 plicae per segment, arranged in rings separated by straight grooves perpendicular to the main body axis. Seven plicae pass to the ventral side between oncopods.

*Mid-line* (also known as dorsomedian furrow): The dorsomedian furrow is conspicuous, forming a channel which splits the folds antero-posteriorly (Fig. 3). Hyaline organs absent along the dorsomedian furrow.

**Jaw:** Outer jaw blade with one principal and one well developed accessory blunt tooth; presence of a vestigial second accessory tooth (Fig. 4A). Inner jaw blade with one principal and two accessory teeth, the first one well developed and blunt and the second less developed or vestigial (Fig. 4B), 9 to 10 denticles (Fig. 4C).

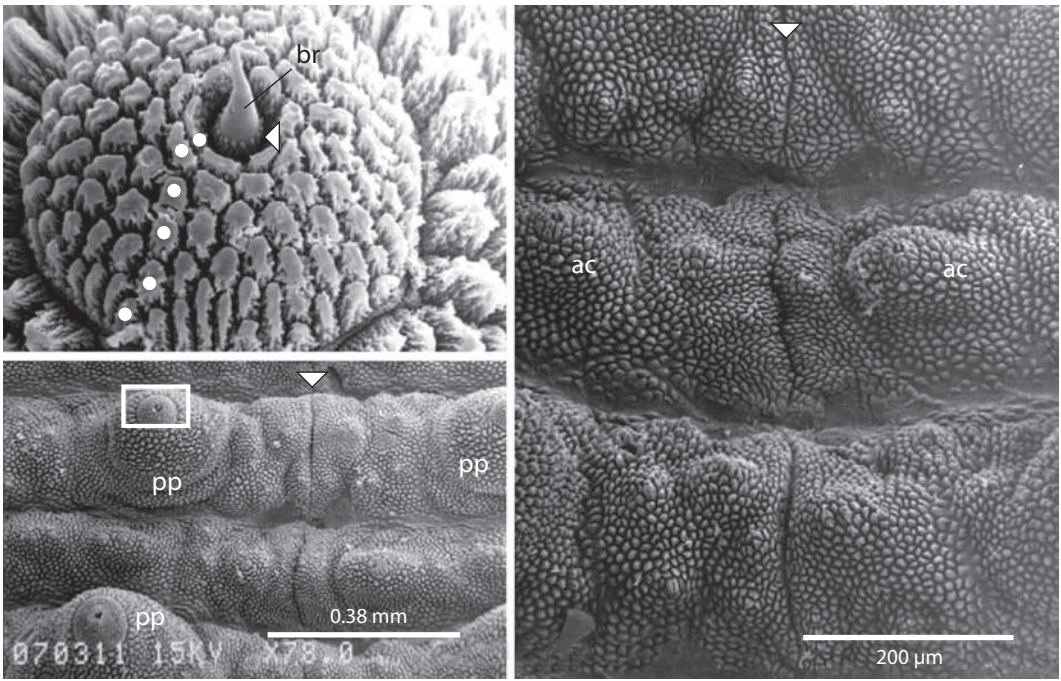

Fig. 3. Upper: apical piece of the primary papillae (pp), with large spherical shape, and a basal diameter of ~20 ranks. The apical piece has 5-6 scale ranks. (◁) The base of the apical seta (sensory bristle) is ornamented. Lower: (▽) the middle dorsal line forms a channel which splits the folds antero-posteriorly (primary papillae). Scale bar= 38 mm. Right: another view of middle dorsal line (▽). Accesory papillae (ac). Scale bar= 200 µm.

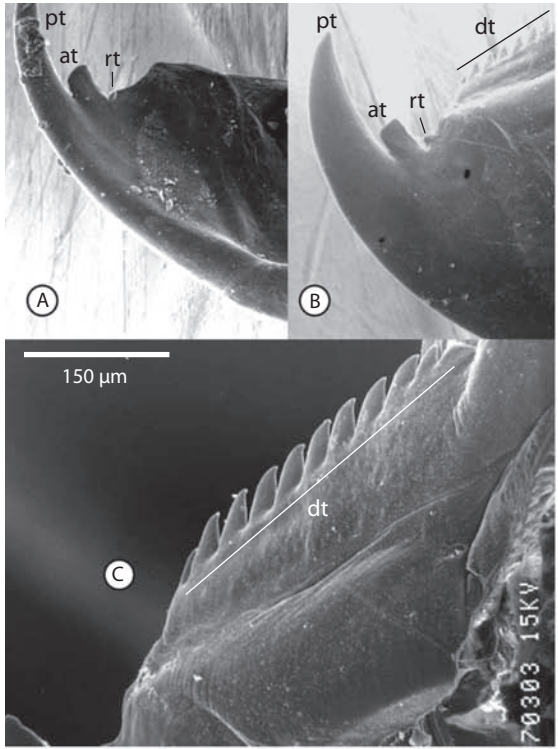
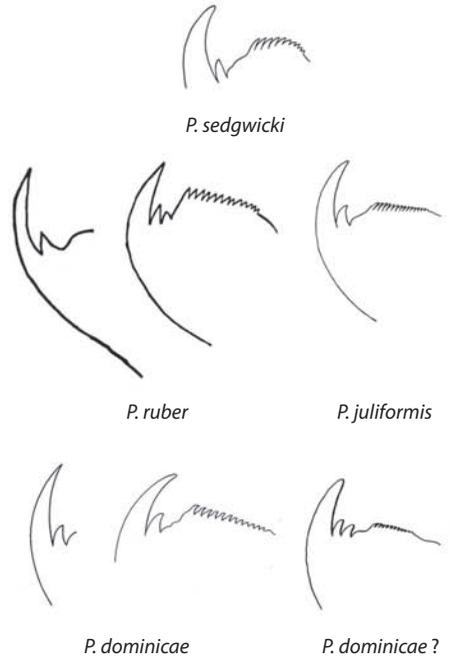
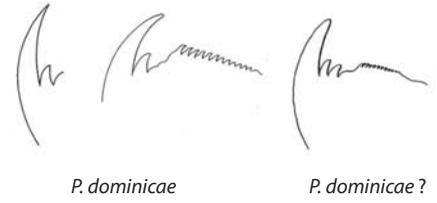
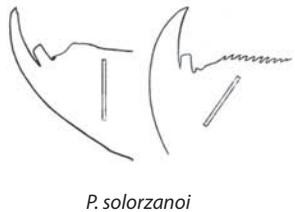
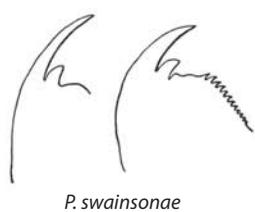
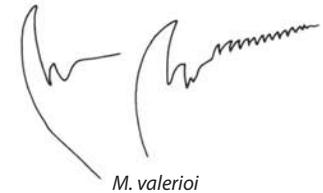
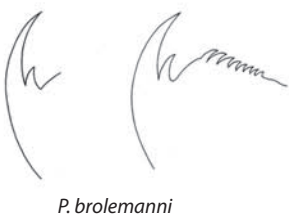
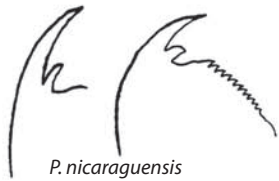
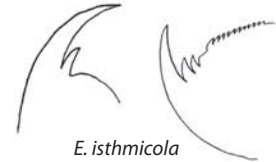
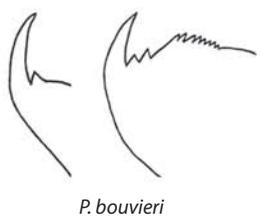
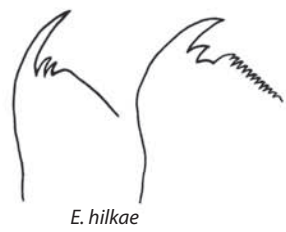
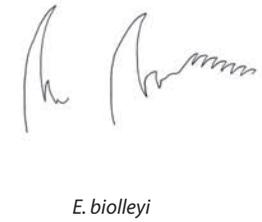

Fig. 4. Jaws. (A) Outer blade of mandible with (pt) one principal tooth, (at) one well developed blunt accessory tooth, and (rt) rudiments of a second accessory tooth. Scale bar= 30mm. (B) Inner blade of mandible with two accessory teeth, the first one blunt and well developed, and a rudimentary second tooth. Scale bar= 30 mm. (C) Inner blade of the mandibule, with 9-10 denticles (dt). Scale bar= 150 $\mu$m.

**Body size:** In life, holotype body length was 22cm including antennae (reduced to 13cm after preservation in 70% alcohol). Other specimens: adults 7.1-11.7cm, young 2.2-4.0cm (in 70% alcohol).

**Color in life:** No dorsal ornamentation but greater dorsal primary papillae may look like dark dots. Oncopods pale or light yellow, contrasting with the darker body. Holotype was light brown (Fig. 5), her newborns were red. Apparently there are two color morphotypes present: light brown or red wine color. We also observed deep brown onychophorans in the area but they might represent a different species.

**Molecular analysis:** The COI DNA sequences of the two *P. solorzanoi*'s specimens differ in 15 bases (2.13%) (Table 1). Identity=644/658 (97%), gaps=0.

Nevertheless, all the differing bases at DNA level do not affect amino acid coding, so the translated aminoacid sequences of 219 length show identical COI proteins in both *P. solorzanoi*'s individuals (Table 2). Such length comparison show a pairwise identity of 97.7% (214 identical sites) between *P. solorzanoi* and *E. biolleyi*, both from the same family, but just 89.5% (196 identical sites) between *P. solorzanoi* and *M. inae*, species from different families. Coherently *E. biolleyi* and *M. inae* show a pairwise identity of 88.6% (194 identical sites).

Note that *E. biolleyi* presents a deletion of 1 amino acid (at shown position no. 50) with respect to the other studied species. Unfortunately the comparison between *P. solorzanoi* and *O. corradoi* is possible just in a length of 151 amino acids, because the studied sequenced fragment is shorter in the last species. They show a pairwise identity of 92.7% (140 identical sites).

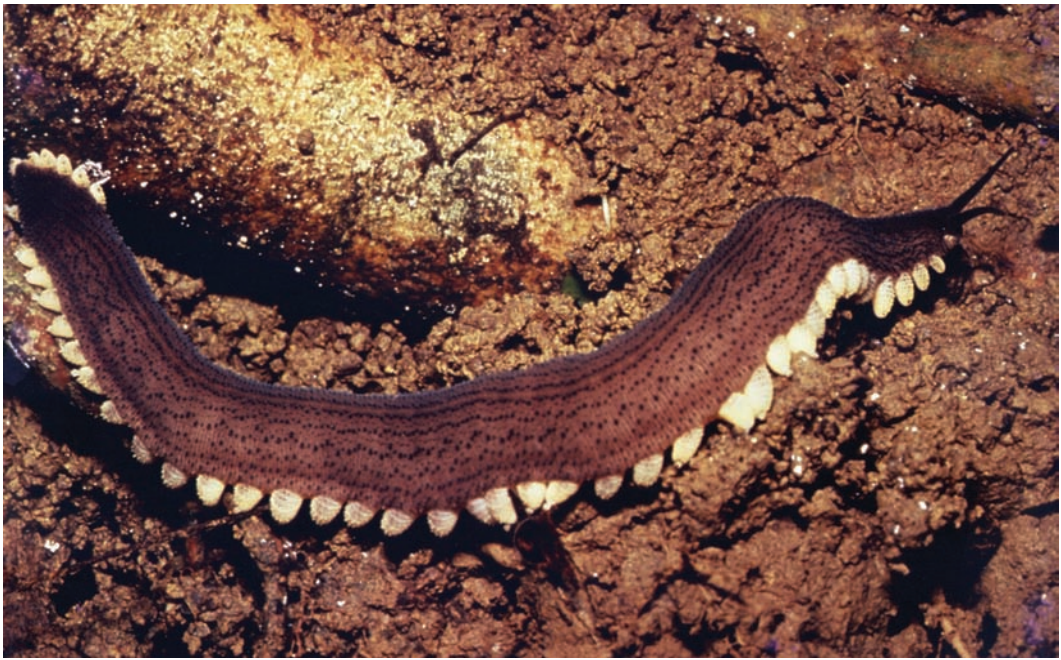

Fig. 5. Brown specimen of *Peripatus solorzanoi* Morera-Brenes and Monge-Nájera 2010.

TABLE. 1

*Nucleotide alignment of two* P. solorzanoi *individuals' sequences of the cytochrome c oxidase subunit I (COI) mitochondrial gene*

```
PE11    1   AACAATATATTTTATTTTTGGTGTATGGTCTGCTATAGTTGGGACTTCGTTAAGTTTATT    60
PE12        AACAATATATTTTATTTTTGGTGTATGGTCTGCTATAGTTGGGACTTCGTTAAGTTTATT

PE11   61   AATTCGAACAGAATTAATAGTTATGGGGAATTTATTAGGTGATGATCAATTGTTTAATGT   120
PE12        AATTCGAACAGAATTAATAGTTATGGGGAATTTATTAGGTGATGATCAATTGTTTAATGT

PE11  121   TATTGTTACGGCTCATGCTTTTGTTATAATTTTTTTTATAGTTATACCAATTATAATTGG   180
PE12        TATTGTTACGGCTCATGCTTTTGTTATAATTTTTTTTTATAGTTATACCAATTATAATTGG

PE11  181   TGGTTTTGGGAATTGGTTAGTTCCTTTAATGTTAGGTGCACCTGATATGGCATTTCCTCG   240
PE12        TGGTTTTGGGAACTGGTTAGTTCCCTTAATGTTAGGTGCGCCTGATATGGCATTTCCTCG

PE11  241   TTTAAATAATTTGAGTTTTTGATTATTACCTCCTTCTTTTTTTTGTTGTTGAGATCTTC   300
PE12        TTTAAATAATTTAAGGTTTTGATTATTACCTCCTTCTTTTTTTTTATTGTTAAGATCTTC

PE11  301   AATGGTTGAAAGAGGAGCCGGAACAGGGTGAACTGTATATCCTCCTTTGTCAAGGAATTT   360
PE12        AATGGTTGAAAGAGGAGCTGGAACAGGGTGAACTGTATATCCTCCTTTGTCAAGGAATTT

PE11  361   GACTCATAGTGGGGGATCTGTTGATTTAACTATTTTTCTTTACATTTGGCTGGTATTTC   420
PE12        AACTCATAGTGGGGGATCTGTTGATTTAACTATTTTTCTTTACATTTGGCTGGTATTTC

PE11  421   ATCAATTTTGGGTGCATTAAATTTTATTACTACTGTGATTAATATGCGAACTTTTGGGAT   480
PE12        ATCAATTTTGGGTGCGTTAAATTTTATTACTACTGTAATTAATATGCGAACTTTTGGGAT

PE11  481   GGTTTTTGAACGTGTTCCTTTATTTGTTTGATCTGTAAAGATCACGGCTATTTTATTGTT   540
PE12        GGTTTTTGAACGTGTTCCTTTATTTGTTTGATCTGTAAAGATTACGGCTATTTTATTGTT

PE11  541   ATTATCGTTACCTGTTTTGGCAGGTGCTATTACTATGTTATTAACTGATCGAAATTTAAA   600
PE12        ATTATCGTTGCCTGTTTTGGCAGGTGCTATTACTATGTTATTAACTGATCGAAATTTAAA

PE11  601   TACTTCATTTTTTGATCCTGCTGGGGGGGGTGATCCTATTTTATATCAACATTTATTT   658
PE12        TACTTCATTTTTTGACCCTGCTGGGGGGGGTGATCCTATTTTATATCAACATTTATTT
```

Individuals: PE11(Holotype), PE12 (Paratype).

**Phylogenetic analysis:** The available information on the COI protein sequence of the species: *P. solorzanoi*, *E. biolleyi*, *O. corradoi* and *M. inae* was used to estimate the genetic distances between such taxa. Genetic distance matrix (substitutions per site) is shown at Table 3.

The phylogenetic analysis at the full comparison 151 amino acids COI fragment is shown at Fig. 6. The Caribbean Peripatidae species *Peripatus* and *Epiperipatus* are close to each other in a node. The Peripatopsidae *M. inae* was used as outgroup.

**Behavior and net weaving:** Individuals of this new species actively forage for prey at night in rivulet banks. They expel adhesive when touched; the amount of adhesive is larger than in other Costa Rican species. They move away from a common flashlight beam. Some individuals were born in captivity, probably prematurely as a result of the mother's post-collecting stress. A high speed film (Fig. 7) shows the animal touching prey with the antennae before expelling two streams of liquid adhesive. Muscular action produces a swinging movement of the adhesive-spelling organs; as

TABLE 2

*Alignment of amino acid sequences of a portion of the mitochondrial cytochrome c oxidase subunit I (COI)*

```
.                     1         10        20        30        40        50        60
                      |         |         |         |         |         |         |
E. biolley      TMYFIFGVWSAMVGTSLSLLIRTELMVMGNLLGDDQLFNVIVTGHAFVM-FFLVMPIMIG
P. solorzanoi   ........................................A.....I..M.......
O. corradoi     ????????????????????????????????????????????????????????????
M. inae         .......S.A........F......TQT...M.....Y..V..A.....I..M.......

E. biolley      GFGNWLVPLMLGAPDMAFPRLNNLSFWLLPPSFFLLLSSSMVESGAGTGWTVYPPLSSNL
P. solorzanoi   ............................................................
O. corradoi     ????????..............M........M..IA........................
M. inae         ......................M........L..IG.......................I

E. biolley      THSGGSVDLTIFSLHLAGISSILGALNFITTVINMRTIGMVFERVSLFVWSVKITAILLL
P. solorzanoi   ................................F.......P..................
O. corradoi     S...A.............V............I.....Y...M...P.............
M. inae         S..............................L....Y..TM..IS..........V...

E. biolley      LSLPVLAGAITMLLTDRNLNTSFFDPAGGGDPILYQHLF
P. solorzanoi   ......................................
O. corradoi     ...........................L.........
M. inae         .A....................................
```

Amino acid sequences of a portion of the mitochondrial cytochrome c oxidase subunit I (COI) inferred from the gene nucleotide sequences in relevant taxa. Dots indicate identity with *E. biolleyi* amino acids used here as reference protein, (-) delection, (?) unstudied segment. No. 1 corresponds to position #14 at *E. biolleyi*. Peripatidae: *E. biolleyi* (ABF93293), *P. solorzanoi* (Code PE11), *O. corradoi* (AAC95414), Peripatopsidae: *M. inae* (ABQ95564).

a result, the streams cross in mid air, weaving the net.

## DISCUSSION

**Placement in the Caribbean Group, genus *Peripatus*:** The new species, *P. solorzanoi* belongs to the Caribbean group, *sensu* Bouvier 1905, by presenting three foot papillae arranged laterally, two anterior and one posterior (Figs. 1, 8) and four complete creeping pads with nephridial tubercle opening between the 3rd and 4th pad (Figs. 8, 9). These features are shared currently by five neotropical genera: *Peripatus* Guilding 1826, *Epiperipatus* (Clark 1913), *Macroperipatus* (Clark 1913), *Plicatoperipatus* (Clark 1913) and *Speleoperipatus* (Peck 1975).

The presence of eyes and only twelve plicae per body segment exclude the possibility of such species being to *Speleoperipatus* or *Plicatoperipatus*, respectively. The rounded bases of dorsal primary papillae exclude also *Macroperipatus*, that bears papillae with quadrangular basis.

TABLE 3

*Genetic distance matrix for the relevant species compared in Fig 1*

|  | *E. biolley* | *M. inae* | *O. corradoi* | *P. solorzanoi* |
|---|---|---|---|---|
| *E. biolley* | - |  |  |  |
| *M. inae* | 0.090226 | - |  |  |
| *O. corradoi* | 0.082974 | 0.090226 | - |  |
| *P. solorzanoi* | 0.013338 | 0.097533 | 0.075777 | - |

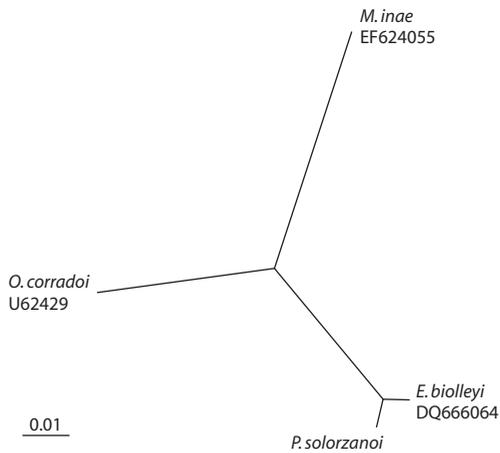

Fig. 6. Neighbor-Joining tree of the Onychophoran Peripatidae: *Epiperipatus biolleyi*, *Peripatus solorzanoi*, *Oroperipatus corradoi* and Peripatopsidae: *Metaperipatus inae* based on Jukes-Cantor's genetic distances.

The distribution of dorsal papillae in the new species does not fit the old definition of *Peripatus* and *Epiperipatus* made by Peck (1975), but the type is similar to *Epiperipatus* (all of one type, Peck 1975). The new species is unique in having 5 to 12 accessory papillae between two of the larger ones. This is somehow similar to *Peripatus* (the primary papillary tubercles separated by rather broad intervals where the accessory papillae occur, Peck 1975). Read (1988a) believed that the distinction between *Peripatus* and *Epiperipatus* was invalid. According to Read's classification (1988a), the number of scale ranks at apical piece in this new species classifies it in the genus *Peripatus* (*sensu strictu*).

**DNA diversity and phylogenetics:** The observed DNA sequence diversity (2.13%) in

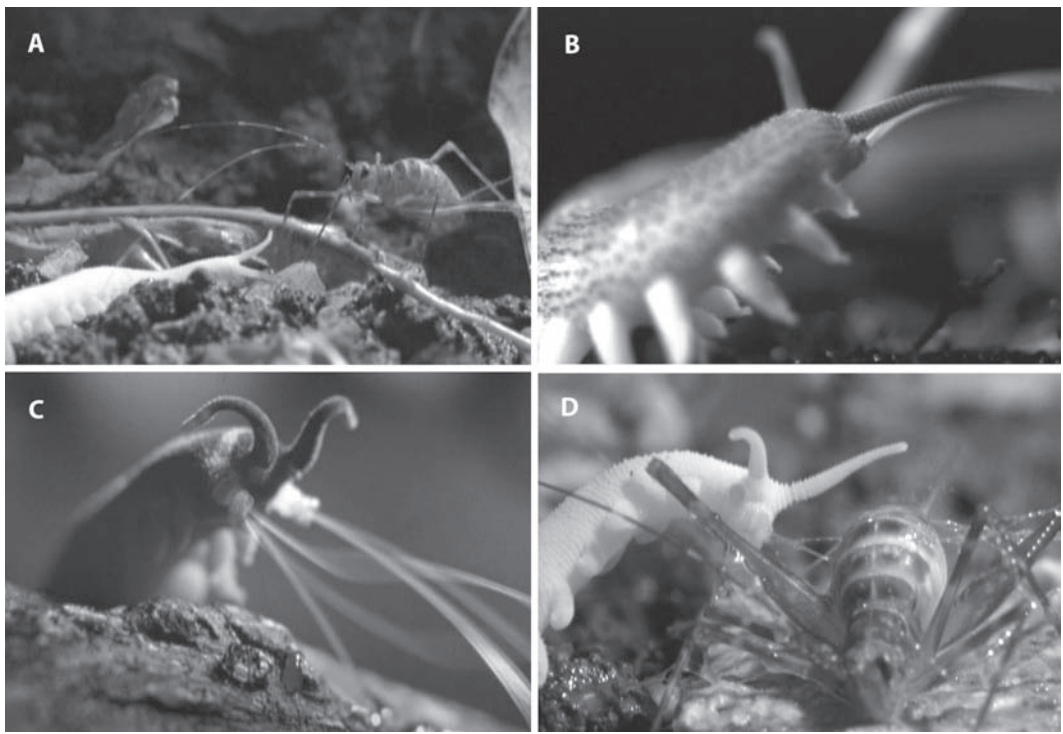

Fig. 7. Prey capture and feeding behavior of the new species: the onychophoran walks at night looking for prey; (A) touches the potential prey with the antennae, and (B) expels two streams of liquid adhesive from specialized organs that (C) oscillate so that the streams cross in mid air and produce the net. D. Finally, when the prey is secured with the adhesive, it is processed with external digestion.

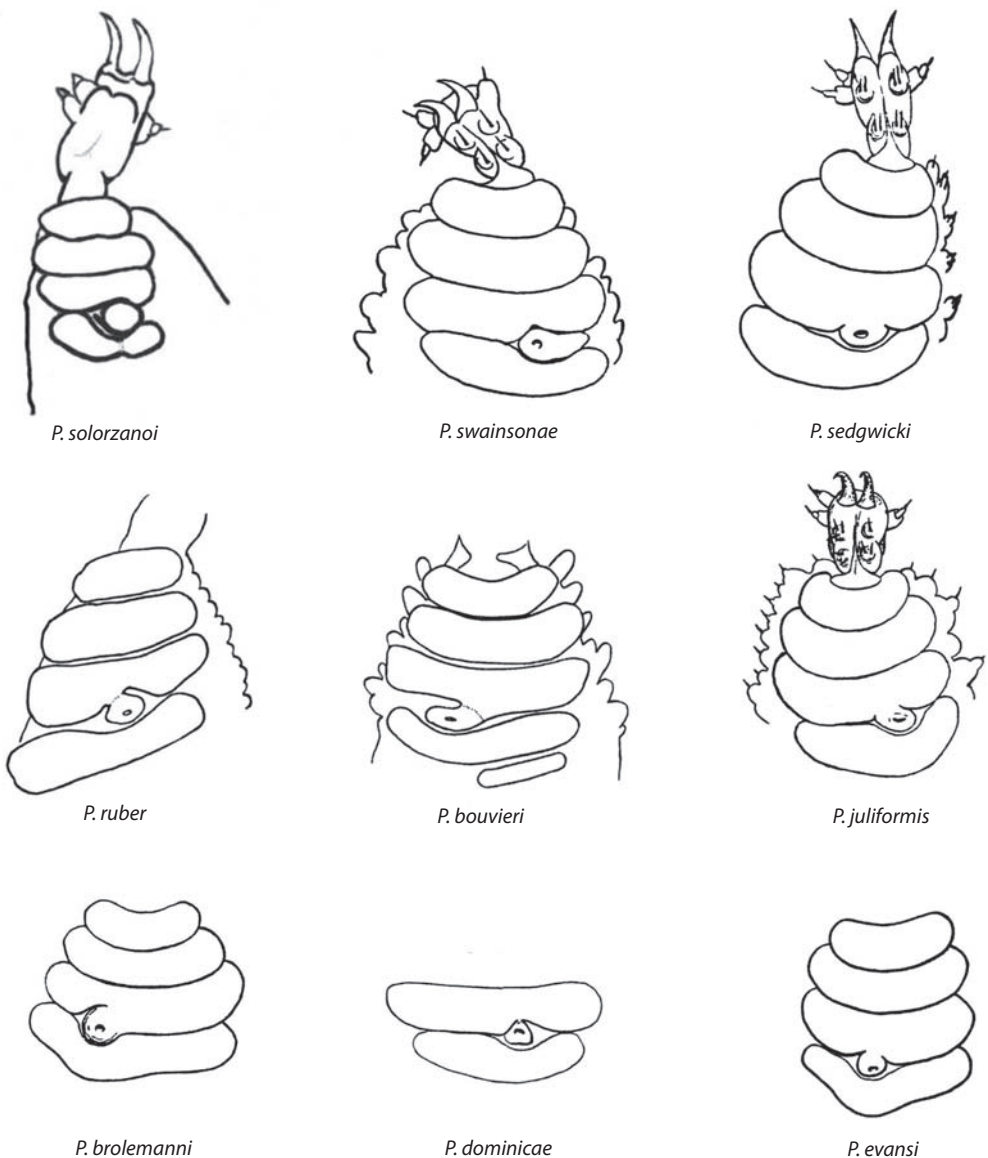

Fig. 8. Oncopod tip of the 4th and 5th leg pairs (ventral view) of the genus *Peripatus* species (confirmed by Scanning Electron Microscopy).

*P. solorzanoi* is a normal value for intraspecific variation, often sister species differ by >8%, while intraspecific variation is up to 5%. In contrast to the DNA variation, the inferred amino acid sequence is conserved in *P. solorzanoi*, suggesting that the majority of amino acid positions are constrained within this region of the COI gene.

The phylogenetic analysis at the COI protein showed that the Central American Caribbean Peripatidae species *P. solorzanoi* and *E. biolleyi* are close to each other, far from the Andean South American *O. corradoi*. The South American Peripatopsidae *M. inae* was used as appropriate outgroup.

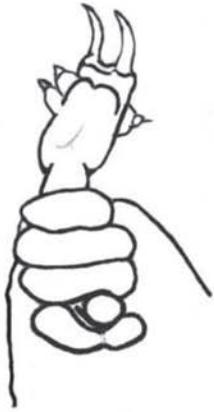
*P. solorzanoi*

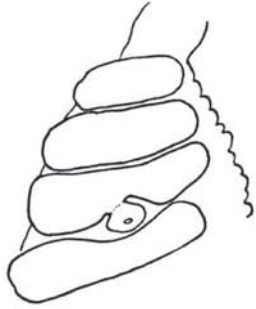
*P. ruber*

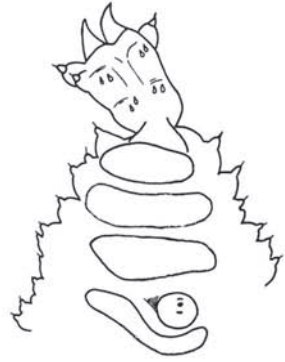
*M. valerioi*

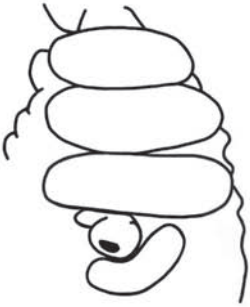
*E. hilkae*

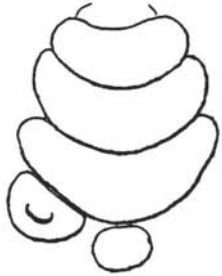
*E. isthmicola*

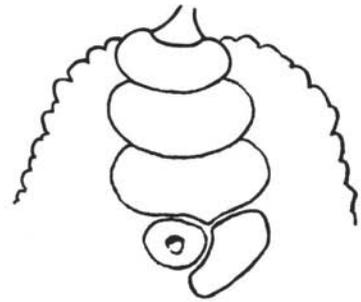
*E. nicaraguensis*

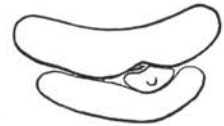
*E. brasiliensis* (from Venezuela)

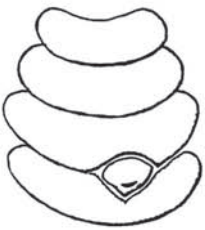
*E. biolleyi*

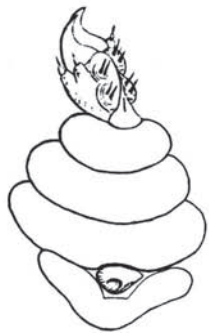
*E. brasiliensis*

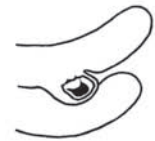
*E. edwardsi* (holotype)

Fig. 9. Oncopod tip of the 4th and 5th leg pairs (ventral view) of the known Costa Rican onychophorans, and *Epiperipatus brasiliensis* and *E. edwardsi*.

**Differences with similar species:** As an aid to identification, we present here some information on how to distinguish the new species from others that occur in the same region. Some of these characteristics may be highly variable within the family, but are reliable within the geographic region considered.

The position of the nephridial tubercle on the fourth and fifth foot distinguishes *P. solorzanoi* from *P. ruber* and *P. bouvieri* because they have "the tubercle largely fused with the third arc" as stated by Bouvier (1905, 1907). Additionally, *P. solorzanoi* does not have the fifth pad that *P. bouvieri* has.

The other species recorded from Costa Rica also differ from *P. solorzanoi*. The new species differs from *E. isthmicola*, *E. hilkae* and *E. nicaraguensis* because its fourth arc is complete. It also differs from *M. valerioi* "where the fourth creeping pad is thin and twists around the urinary tubercle, which is completely free and outerly bound" (Morera-Brenes & León 1986). If the size is not considered, the new species resembles *Epiperipatus biolleyi* (Fig. 9), but it can be separated because the third pad is not indented in *P. solorzanoi*.

The mandible's outer blade formula in *P. solorzanoi* (1.1.1) is different from all confirmed and unconfirmed species of *Peripatus* (as detailed by Read 1988a, b); and it is also different from the other Costa Rican onychophorans (Morera-Brenes & León 1986). Only *Plicatoperipatus* from Jamaica has a similar outer blade formula. The outer blade shape resembles that of *Macroperipatus torquatus* from Trinidad, except that the accessory teeth have a more pointing shape in *M. torquatus* (Read 1988a, b).

The inner blade formula of *P. solorzanoi* (1/1/1(9-10) is similar to that of *Macroperipatus valerioi*: (1/1/1/11) (Morera-Brenes & León 1986). The shape of the inner blade also resembles that of *M. valerioi*, except that in *M. valerioi* has more denticles and the accessory tooth is more acute (Morera-Brenes & León 1986). It distantly resembles the accessory tooth in the inner blade of "*E. brasiliensis*" from Venezuela (in quotation marks because we believe this is a misidentification). However, both *M. valerioi* and "*E. brasiliensis*" have a well developed second accessory tooth that is lacking in the new species. The bilobular shape of the inner blade diastema resembles those of *P. sedgwicki* from Venezuela and *P. swainsonae* from Jamaica.

Both males and females of the new species have more oncopods than other Neotropical *Peripatus* (*sensu stricto*) and other Costa Rican onychophorans (*Epiperipatus* and *Macroperipatus*).

**Geographic distribution:** These are the species of *Peripatus* that have been corroborated with SEM: *P. juliformis* Guilding from St. Vincent, *P. d. dominicae* Pollard from Dominica, *P. antiguensis* Bouv. from Antigua and *P. antiguensis* from Montserrat, *P. d. lachauxensis* Brues from Haiti, *P. swainsonae* Cockerel from Jamaica, *P. sedgwicki* Bouv. from Venezuela, and the new *P. solorzanoi* from Costa Rica (Fig. 10). When only these species are considered, *Peripatus* has a circum-Caribbean distribution, reaching northward from South America into the Central American bridge and Antillean arc. All the Antillean species occur on continental islands, so we hypothesize that the genus colonized the current islands in times of a lower sea level, as suggested by Monge-Nájera (1996). The exception to this hypothesis was Barbados, an oceanic island reported to have a *Peripatus* species, but Read (1988a, b) concluded that it was misidentified and relocated the species in another genus (as *Epiperipatus barbadensis*).

The unconfirmed continental species of *Peripatus* (*P. ruber* from Costa Rica, *P. bouvieri* from Colombia, *P. brolemani* from Venezuela, *P. torrealbai* from Venezuela (Scorza 1953), *P. bavaysi* from Guadalupe, *P. daniscus* from St. Tomas, *P. daniscus juanensis* from Porto Rico, *P. manni* from Haiti and *P. haitienisis* Brues from Haiti) fit our hypothesis. If they were removed from *Peripatus* in the future, it would not affect our hypothesis.

In contrast with *Peripatus*, the genus *Epiperipatus* seems to be more widely distributed in South America. Nevertheless, both genera have the same circum-Caribbean distribution, suggesting they originated from a common ancestor before their parallel expansion (Monge-Nájera 1995). We must note, however, that maybe *Peripatus* is not a monophyletic genus.

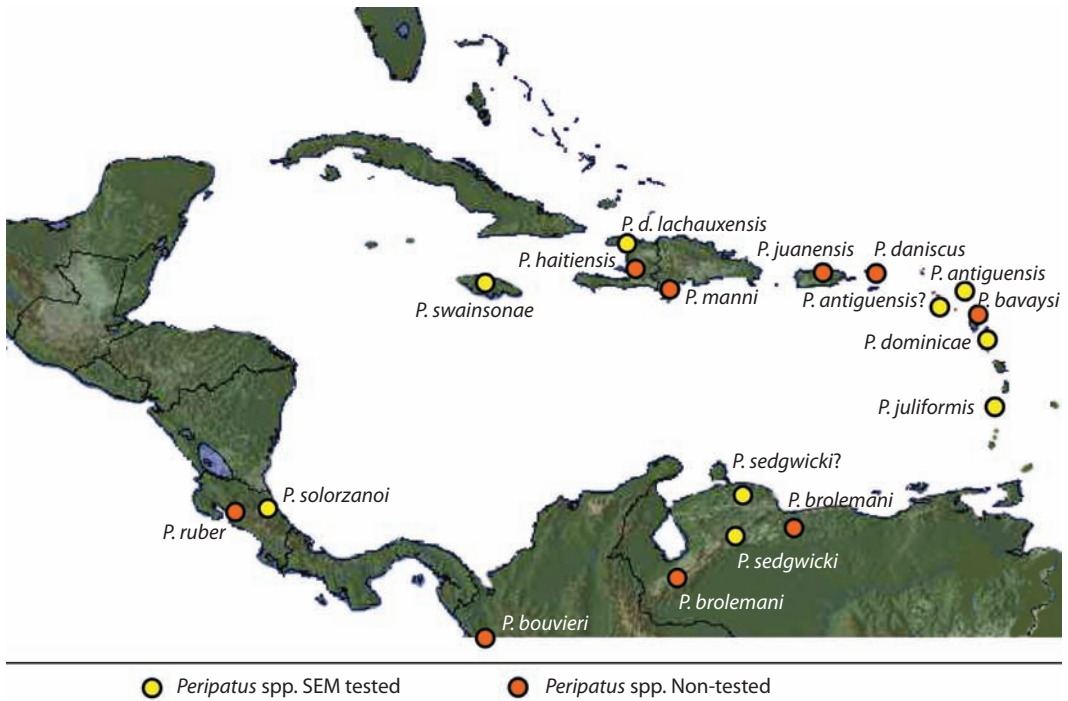

Fig. 10. Geographic distribution of *Peripatus*. Open circles (yellow): species of *Peripatus* that have been confirmed with Scanning Electron Microscopy. Closed circles (orange): other described species of *Peripatus*.

**Body size:** The body size has been regarded as of little taxonomic use due to the contraction capacity of Onychophora. Thus, it is difficult to obtain a precise measurement of body length, mainly after fixation, when several degrees of contraction may occur (Read 1988a). Despite Read's statement (1988a), the body size of *P. solorzanoi* is quite uncommon among extant and also extinct onychophorans and should be regarded as a diagnostic feature in addition with other morphological characters. The size has been already used to help the diagnosis of *Metaperipatus inae* Mayer 2007 and is very useful to characterize adults of *P. solorzanoi*.

Read (1988a) has previously commented: "The largest onychophorans include the large species of *Oroperipatus* (section 1 of Peripates Andicoles in Bouvier 1905), and *Macroperipatus torquatus*. Certain *Epiperipatus*, including *E. lewisi*, reach a large size (Arnett 1961)". Of these, *Macroperipatus torquatus* was, so far, the largest living species with a total length of 15cm (Read 1985). The new species, *P. solorzanoi*, is even longer with 22cm and its size can be only compared with extinct taxa, namely as the fossil *Xenusion* Pompeckj, 1927 from the Baltic, with 20cm (Dzik & Krumbiegel 1989) and the not fully identified fossil *Jianshanopodia decora* that may reach an equal size if confirmed as being an onychophoran (Liu *et al.* 2006).

Based on a morphometric analysis, Monge-Nájera & Morera (1994) suggested that the number of oncopods is simply a function of the animal's length, but the correlation has important exceptions.

The Onychophora's ancestors were a few millimeters long, lived under water and had few but long oncopods (Monge-Nájera 1994, 1995), so we suspect that at the extreme size of 22cm, *P. solorzanoi* is close to the practical limit for a functional onychophoran because they lack a hard skeleton and cannot excavate their own burrows (Monge-Nájera *et al*. 1993).

**Conservation:** The Onychophora's conservation has been previously discussed (Mesibov 1990, New 1995) and as all other species of Onychophora, *P. solorzanoi* is rare. Despite additional effort to find the species during fieldwork in the Caribbean of Costa Rica, it has never been found outside the limited area where the type was collected. Furthermore, much of the original tropical rainforest in that area has been deforested for farming and housing development. So far, the only known population is limited to a few kilometers of riparian vegetation. So the species is at least under the *Vulnerable* IUCN category (IUCN 2000). Under these circumstances, we believe that active protection of the habitat of the largest onychophoran ever described, is urgent.

ACKNOWLEDGMENTS

We appreciate the assistance and support of Alejandro Solórzano, who brought the new species to our attention and provided the photograph in Fig. 5-B, José A. Vargas Z. for the photograph in Fig. 5-A, and the British Broadcasting Corporation, through Tim Haynes, for allowing us to reproduce frames from the high speed video of *P. solorzanoi* (Fig. 7). We are indebted to Lars Podsiadlowski for his great support with DNA sequencing and critical reading of the paper. Francisco Hernández helped greatly with the SEM. We also thank Hilke Ruhberg, Randall Rubí Chacón, Robert Mesibov, V.M.S.J. Read and three anonymous reviewers for suggestions to improve an earlier draft, as well as Sergio Aguilar for preparing the final version of figures.

RESUMEN

Los onicóforos o "peripatos" son animales escasos y poco conocidos. Aquí informamos el descubrimiento del onicóforo más grandes conocido, una hembra de 22cm de longitud del bosque costero caribeño de Costa Rica. Analizamos los especímenes con microscopia electrónica de barrido. La nueva especie, *Peripatus solorzanoi*, **sp. nov.** se caracteriza así: papilas primarias convexas y cónicas, con bases redondeadas y más de 18 filas de escamas de alto. Sección apical grande y esférica, con diámetro basal de al menos 20 filas. Parte apical con 6-7 filas de escamas de alto. Fórmula de lámina mandibular externa 1/1/1, lámina interna: 1/1/1/9-10. Diente accesorio romo en ambas láminas. Cuatro almohadillas en cuarto y quinto oncopodios; cuarta almohadilla arqueada. Describimos el mecanismo, previamente desconocido, mediante el cual tejen su red los onicóforos, usando fotografías para mostrar que es de naturaleza muscular. Como todos los demás onicóforos, *P. solorzanoi* es una especie escasa: consideramos urgente la protección del hábitat del onicóforo más grande del mundo.

**Palabras clave:** especie nueva, mecanismo de tejido de red, onicóforo más grande, onicóforos fósiles y vivientes.